\documentclass[11pt,a4paper]{article}

\usepackage{fullpage}
\usepackage{amsmath,amssymb}
\usepackage{hyperref}
\usepackage{graphicx}

\begin{document}

\renewcommand{\topfraction}{1} 
\renewcommand{\bottomfraction}{1}
\renewcommand{\floatpagefraction}{1}
\renewcommand{\textfraction}{0}


\newcommand{\ie}{i.e.,\ }
\newcommand{\eg}{e.g.,\ }

\newcommand{\const}{\operatorname{const.}} 


\newcommand{\rmd}{\,\mathrm{d}}

\newcommand{\Tr}{\operatorname{tr}}

\newcommand{\e}[1]{\operatorname{e}^{#1}}

\newcommand{\op}{\mathcal{O}}

\newcommand{\vev}[1]{\left\langle #1 \right\rangle}
\newcommand{\fvev}[1]{\langle #1 \rangle}
\newcommand{\jvev}[1]{\left\langle j\left| #1 \right| j \right\rangle}

\newcommand{\comb}[2]{\begin{pmatrix} #1\\#2\end{pmatrix}} 

\newcommand{\Order}{\mathcal{O}}

\newcommand{\Mpl}{M_{\text{P}}}
\newcommand{\Lpl}{L_{\text{P}}}

\newcommand{\hypF}{\operatorname{F}}

\newcommand{\iko}{\frac{ik}{\omega}}
\newcommand{\ikto}{\frac{ik}{2\omega}}

\newcommand{\bx}{\mathbf{x}}

\begin{center}

{\Large \textbf{Quantum Portrait of a Black Hole with P\"oschl-Teller Potential}}\\[2em]

\renewcommand{\thefootnote}{\fnsymbol{footnote}}
Wolfgang M{\"u}ck${}^{a,b}$\footnote[1]{wolfgang.mueck@na.infn.it} and Giancarlo Pozzo$^a$\footnote[2]{giancarlo.poz@gmail.com}\\[2em]
\renewcommand{\thefootnote}{\arabic{footnote}}
${}^a$\emph{Dipartimento di Fisica, Universit\`a degli Studi di Napoli "Federico II"\\ Via Cintia, 80126 Napoli, Italy}\\[1em] 
${}^b$\emph{Istituto Nazionale di Fisica Nucleare, Sezione di Napoli\\ Via Cintia, 80126 Napoli, Italy}\\[2em]

\abstract{We improve upon the simple model studied by Casadio and Orlandi [JHEP 1308 (2013) 025] for a black hole as a condensate of gravitons. Instead of the harmonic oscillator potential, the P\"oschl-Teller potential is used, which allows for a continuum of scattering states. The quantum mechanical model is embedded into a relativistic wave equation for a complex Klein-Gordon field, and the charge of the field is interpreted as the gravitational charge (mass) carried by the graviton condensate.}

\end{center}

\section{Introduction}
\label{intro}

In a recent series of papers, Dvali and Gomez have proposed and studied what they call the quantum $N$-portrait of a black hole \cite{Dvali:2011aa, Dvali:2012gb, Dvali:2012rt, Dvali:2012wq, Dvali:2012en}. According to their proposal, black holes are nothing but a system of a very large number, $N$, of soft gravitons at the verge of a quantum phase transition to a self-sustained Bose-Einstein condensate (BEC). Hawking radiation and black hole evaporation are a consequence of condensate depletion, which is due to graviton-graviton interactions, while the presence of an event horizon in the classical metric description is understood as an artefact of the large-$N$ semi-classical limit, in which the quantum $1/N$ hair has been neglected. This latter statement is qualitatively similar to results from the fuzzball proposal, see \cite{Mathur:2005zp} for a review and \cite{Mathur:2014nja} for a more recent contribution. Both proposals state that the classical description of gravity breaks down at the horizon, not just at the singularity, as has been the standard point of view. The quantum $N$-portrait may also shed light upon the species problem \cite{Dvali:2012uq}, whereas extending the idea of quantum compositeness to other gravitational backgrounds has far-reaching consequences for cosmology \cite{Binetruy:2012kx, Dvali:2013eja}, for earlier ideas see, \eg \cite{Fukuyama:2006gp}. For relates studies, see \cite{Flassig:2012re, Dvali:2013lva, Dvali:2013vxa}.

The quantum-$N$ portrait is a logical consequence of another recent idea, namely classicalization of certain field theories, which are non-renormalizable in the Wilsonian sense \cite{Dvali:2010jz, Dvali:2010bf, Dvali:2011th, Alberte:2012is, Kovner:2012yi, Vikman:2012bx, Berkhahn:2013woa}. The crucial statement here is that such field theories achieve ultra-violet completeness by the presence of classical field configurations called \emph{classicalons}, which prevent ultra-short distances from being probed. Classicalons are generically characterized by a very large expectation number, $N$, of field quanta. The classicalization phenomenon is most efficient in gravity, where 
black holes play the role of the classicalons, with an effective graviton number given by
\begin{equation}
\label{intro:N}
	N= \frac{M^2}{\Mpl^2}~,
\end{equation}
where $M$ is the black hole mass and $\Mpl$ is the Planck mass. Here and in what follows, possible numerical factors of order unity are dropped. As a consequence of \eqref{intro:N}, the mean energy of a single graviton is $M/N = \hbar \omega$, and their effective de~Broglie wavelength is of about the size of the gravitational radius, $1/\omega=MG$.\footnote{Planck's constant and the gravitational constant are expressed in terms of the Planck mass and length by $\hbar=\Mpl\Lpl$ and $G=\Lpl/\Mpl$, respectively.}
In fact, \eqref{intro:N} has a much more general meaning. As argued in \cite{Mueck:2013mha, Mueck:2013wba} in analogy with electromagnetism, each macroscopic body of mass $M$ generates a coherent state of gravitons, with the occupation number of graviton quanta given by \eqref{intro:N}. The energy carried by the gravitational field is such that the average de~Broglie wavelength of the gravitons is of about the size of the body's geometric dimensions. Graviton condensation, which corresponds to black hole formation, occurs when the gravitational energy coincides with the total energy (Komar mass).  

In the absence of a quantum theory of gravity, toy models are useful tools to test ideas such as the quantum-$N$ portrait. 
The harmonic black hole model of Casadio and Orlandi \cite{Casadio:2013hja, Casadio:2013iqc} describes the graviton BEC as an anisotropic fluid with the peculiar equation of state $p_\parallel=-\varepsilon$ (see Section~\ref{bh} for the notation), and with an energy density $\varepsilon$ proportional to the particle density in the ground state of a three-dimensional harmonic oscillator. With reasonable choices of the model parameters, they showed that the solution of Einstein's equation describes indeed a regular black hole, \ie a gravitational configuration with an horizon, but without a singularity. 

In the present paper, we improve upon Casadio and Orlandi's work in the following respects. 
Instead of the harmonic oscillator, which has only bound states, we shall consider the P\"oschl-Teller potential \cite{Poeschl}, which has a continuum above the bound states. The P\"oschl-Teller potential is one of only a handful of potentials for which the Schr\"odinger equation is analytically solvable. This makes it quite unique as a toy model. Moreover, it is interesting to note that the P\"oschl-Teller potential has appeared in a recent study \cite{Husain:2012im} on the \emph{generalized uncertainty principle}, which is another approach to reconcile gravity with quantum physics \cite{Kempf:1994su, Kempf:1996nk, Isi:2013cxa}.
A non-relativistic model containing continuum states, however, is inconsistent, because most of the scattering modes would propagate with superluminal velocities. Therefore, we construct a relativistic model of a complex Klein-Gordon field in an external potential, the solutions of which can be found from the non-relativistic model. There are a number of very interesting and useful implications of our relativistic embedding. First, the number of a priori free parameters in the model is reduced to a single scale parameter, $\omega$, which we identify as the effective de~Broglie frequency of the gravitons. Second, the charge density of the Klein-Gordon field is interpreted as the gravitational charge density (\ie mass density) of the graviton condensate, which enters in Einstein's equation as the energy density $\varepsilon$ of the anisotropic fluid. 
In summary, our model differs from \cite{Casadio:2013hja, Casadio:2013iqc} in the use of the P\"oschl-Teller potential and a relativistic wave equation. In other respects we shall stick to their ideas, in particular to the marginal binding condition, which determines the strength of the potential, and the use of the anisotropic fluid energy momentum tensor with $p_\parallel=-\varepsilon$. Our results confirm that for the most natural value of $\omega$ the gravitational solution is a regular black hole.

The rest of the paper is structured as follows. In Section~\ref{qm}, the $l=0$ modes of the nonrelativistic Schr\"odinger equation with a spherically symmetric P\"oschl-Teller potential are found. Modes with $l>0$ are not considered, because it is implicitly assumed that the potential is somehow self-generated by the BEC, so that the assumption of spherical symmetry would not be self-consistent in the presence of angular momentum. The relativistic model involving a complex Klein-Gordon field is constructed in Section~\ref{rel} in such a way that the solutions found in Section~\ref{qm} can be readily used. In Section~\ref{bh}, the gravitational solution of the toy model is found. Finally, Section~\ref{conc} contains the conclusions.

\section{Quantum mechanical toy model}
\label{qm}

The quantum mechanical toy model we are considering is described by the time-independent Schr{\"o}dinger equation for a particle of mass $m$ moving in a spherically symmetric P{\"o}schl-Teller potential
\begin{equation}
\label{qm:schrodinger}
	\left[ -\frac{\hbar^2}{2m} \, \nabla^2 +V(r) \right] \Psi = E \Psi~,
\end{equation}
where \cite{Poeschl}
\begin{equation}
\label{qm:poschl}
	V(r)=-\frac{\hbar^2\omega^2\lambda(\lambda+1)}{2m\cosh^2(\omega r)}~.
\end{equation}
Written in this form, $\omega$ is a constant of dimension $L^{-1}$ defining the characteristic length scale of the system, and $\lambda>0$ is a dimensionless parameter, which we shall fix later.\footnote{We have shifted $\lambda$ by $1$ with respect to \cite{Poeschl,Flugge}.} 

The standard decomposition of the wave function into spherical harmonics,
\begin{equation}
\label{qm:Psi.decomp}
	\Psi = \frac{\chi(r)}r Y_{lm}(\Omega)
\end{equation}
gives rise to the radial Schr{\"o}dinger equation
\begin{equation}
\label{qm:schroedinger.radial}
	\left[ -\frac{\hbar^2}{2 m} \partial_r^2 + \frac{\hbar^2 l(l+1)}{2 m r^2} + V(r) 
	\right] \chi(r) = E \chi(r)~,
\end{equation}
with the condition 
\begin{equation}
\label{qm:r0cond}
	\chi(r)=\Order(r)\qquad \text{for $r\to 0$.} 
\end{equation} 

We will restrict our attention to the $l=0$ modes, because in a self-consistent model there would be no reason to assume the spherical symmetry of the potential for non-zero angular momentum. Thus, \eqref{qm:schroedinger.radial} reduces to a one-dimensional Scr\"odinger equation with potential $V(r)$. Because of \eqref{qm:r0cond}, only the odd eigenfunctions are allowed. For the P{\"o}schl-Teller potential, the solutions are known analytically. Let us briefly summarize the solution of \eqref{qm:schroedinger.radial} following the presentation of \cite{Flugge}. 

After defining a wave number $k$ by 
\begin{equation}
\label{qm:k.def}
	E = \frac{\hbar^2 k^2}{2m}~,
\end{equation}
changing variable to 
\begin{equation}
\label{qm:rho.def}
	\rho = -\sinh^2(\omega r)~,
\end{equation}
which implies $\rho \in(-\infty,0)$, and writing 
\begin{equation}
\label{qm:f.def}
	\chi(r) = (1-\rho)^{-\lambda/2} f(\rho)~,
\end{equation}
\eqref{qm:schroedinger.radial} gives rise to a hypergeometric differential equation for the function $f(\rho)$,
\begin{equation}
\label{qm:hyp.geom}
	\left\{ \rho(1-\rho) \partial_\rho^2 +\left[\frac12-(1-\lambda) \rho\right] \partial_\rho 
	-\frac14 \left(\frac{k^2}{\omega ^2}+\lambda ^2\right) \right\} f(\rho) = 0~.
\end{equation}
Then, with the parameters
\begin{equation}
\label{qm:ab.def}
	a = \frac12 \left(- \lambda - i \frac{k}{\omega} \right)~, \qquad 
	b = \frac12 \left(- \lambda + i \frac{k}{\omega} \right)~,
\end{equation}
the general solution of \eqref{qm:hyp.geom} is 
\begin{equation}
\label{qm:hyp.geom.sol}
	f(\rho) = a_1 \hypF \left(a,b;\frac12;\rho \right) 
	+ a_2 \rho^\frac12 \hypF \left(a+\frac12,b+\frac12;\frac32;\rho \right)~,
\end{equation}
where $\hypF\equiv{}_2\mathrm{F}_1$ denotes the Gauss hypergeometric function \cite{Gradshteyn}, and $a_1$ and $a_2$ are constants. The condition \eqref{qm:r0cond} requires $a_1=0$. Therefore, the solution for $\chi(r)$ is found to be
\begin{equation}
\label{qm:chi.sol}
	\chi(r) = C \frac{\sinh(\omega r)}{\cosh^\lambda(\omega r)} 
	\hypF \left(a+\frac12,b+\frac12;\frac32;-\sinh^2(\omega r) \right)~,
\end{equation}
with $C$ denoting a normalization constant. 

To find the asymptotic behaviour of $\chi$ for large $r$, one can use a transformation formula such as 9.132 of \cite{Gradshteyn}, with the result
\begin{equation}
\label{qm:chi.asympt}
	\chi(r) \approx \frac14 C \left( c_1 \e{ikr} + c_2 \e{-ikr} \right) \qquad (r\to \infty)~,
\end{equation}
where $c_1$ and $c_2$ are the two constants
\begin{equation}
\label{qm:c12}
	c_1 = \frac{\Gamma\left(\ikto\right) \Gamma\left(\frac12+\ikto\right)}{%
	\Gamma\left(\frac{1-\lambda}2+\ikto \right) \Gamma\left(\frac{2+\lambda}2+\ikto \right)}~, \qquad 
	c_2 = \frac{\Gamma\left(-\ikto\right) \Gamma\left(\frac12-\ikto\right)}{%
	\Gamma\left( \frac{1-\lambda}2-\ikto \right) \Gamma\left(\frac{2+\lambda}2-\ikto \right)}~.
\end{equation}
For $E>0$, \eqref{qm:chi.asympt} describes a superposition of outgoing and incoming spherical waves. These solutions form the continuum of scattering states. 
For $E< 0$, it is useful to define
\begin{equation}
\label{qm:kappa.def}
	k= i\kappa~, \qquad (\kappa\geq 0)~.
\end{equation}
The normalizability of the wave function for bound states requires $c_2=0$, which implies that $\kappa$ must assume discrete values,
\begin{equation}
\label{qm:spectrum}
	\frac{\kappa_n}{\omega} = \lambda - 1 -2n~, \qquad n=0,1,\ldots, \left[\frac{\lambda-1}2 \right]~.
\end{equation}

In a non-relativistic setting, the above analysis entirely solves the problem of finding the eigenstates and the spectrum of the Hamiltonian of our toy model, for any values of the parameters $m$, $\omega$ and $\lambda$. Moreover, the dynamics of the quantum states would be given in terms of the time-dependent Schr\"odinger equation. It is easy to see that this would lead to superluminal scattering modes, which impedes a physical interpretation. To obviate this problem, we need to embed our non-relativistic toy model into a relativistic wave equation. This will be done in the next section. As a very welcome by-product of this embedding, the mass parameter $m$ will cease to be an independent quantity, and the microscopic toy model will be characterized solely by the inverse length scale $\omega$.

\section{Relativistic toy model}
\label{rel}

Let $\Psi$ now be a complex Klein-Gordon field. The Klein-Gordon equation admits the introduction of two independent potentials, 
\begin{equation}
\label{rel:KG}
	\left\{ -\left[i\hbar\partial_t -V(\bx)\right]^2 -\hbar^2 \nabla^2 + \left[S(\bx) + \mu\right]^2 \right\} \Psi(t,\bx) =0~,
\end{equation}
where $V(\bx)$ and $S(\bx)$ are the ``vector'' and ``scalar'' potentials, respectively, and $\mu$ is the rest mass.\footnote{$V$ behaves as the time component of the electromagnetic vector potential, $S$ transforms as a scalar under Lorentz transformations.} For time-independent potentials, we make the ansatz 
\begin{equation}
\label{rel:psi.t}
	\Psi(t,\bx) = \e{-i\epsilon t/\hbar} \Psi(\bx)~,
\end{equation}
so that $i\hbar\partial_t$ in \eqref{rel:KG} can be replaced by the energy $\epsilon$. We shall be interested in the positive-energy modes only. If one adopts the particular choice $S=V$,\footnote{Interest in this case has arisen recently in the context of the physics of nuclei and is motivated by the pseudospin symmetry. It is intriguing that, just as in the free case, for $S=\pm V$ the Klein-Gordon equation is the square of a Dirac equation. See \cite{Alhaidari:2005fw} and references therein.} \eqref{rel:KG} gives rise to
\begin{equation}
\label{rel:schrodinger}
	\left[ -\frac{\hbar^2}{2(\epsilon+\mu)} \nabla^2 +V - \frac12(\epsilon - \mu) \right] \Psi(\bx) = 0~,
\end{equation}
which is just \eqref{qm:schrodinger} with 
\begin{equation}
\label{rel:coeffs}
	m= \epsilon+\mu~, \qquad E= \frac12 \left(\epsilon-\mu\right)~.
\end{equation}
Therefore, the time-independent Schr\"odinger equation \eqref{qm:schrodinger} and its solutions can be straightforwardly promoted to a relativistic toy model. In this setting, $m$ ceases to be an independent mass parameter. Combining \eqref{rel:coeffs} with \eqref{qm:k.def} yields the relativistic dispersion relation 
\begin{equation}
\label{rel:dispersion}
	\epsilon^2 = \mu^2 +\hbar^2 k^2~.
\end{equation}
Because the potential $V$ in \eqref{rel:schrodinger} must be independent of $\epsilon$, the parameter $\lambda$ in  \eqref{qm:poschl} becomes energy-dependent. Let us denote this by a subscript. Thus, let us write
\begin{equation}
\label{rel:lambda.E}
	\lambda_\epsilon (\lambda_\epsilon+1) = \frac{\epsilon +\mu}{\mu} \xi~,
\end{equation}
with a constant $\xi$ that will be determined presently. 

To fix the parameters, we use the same conditions as in \cite{Casadio:2013hja}, namely the existence of a single bound state, which is the ground state, and the coincidence of the would-be first excited bound state with the onset of the continuum (marginal binding). The marginal binding condition implements the statement that quanta excited by condensate depletion must be emitted from the black hole in the form of Hawking radiation. In addition, we shall minimize the energy of the ground state by setting $\epsilon_0=0$,\footnote{Remember that we do not consider the negative energy solutions.} which makes the ground state wave function time-independent. 

Therefore, using \eqref{qm:spectrum} with $n=1$, the marginal binding condition determines the value of $\lambda$ at the onset of the continuum,
\begin{equation}
\label{rel:lambda.1}
	\lambda_{k=0} = 3~.
\end{equation}
Then, because of $\epsilon_{k=0}=\mu$ from \eqref{rel:dispersion}, \eqref{rel:lambda.1} and \eqref{rel:lambda.E} yield  
\begin{equation}
\label{rel:xi}
	\xi=6~.
\end{equation}
Furthermore, $\epsilon_0=0$ implies that the effective graviton mass $\mu$ coincides with the energy gap above the ground state. To find the value of $\lambda$ for the ground state, we use again \eqref{rel:lambda.E}, which gives
\begin{equation}
\label{rel:lambda.0}
	\lambda_0 = 2~.
\end{equation}
Substituting this into \eqref{qm:spectrum} for $n=0$, one finds $\kappa_0=\omega$. In turn, this implies with \eqref{rel:dispersion} that also $\mu$ is not an independent parameter, but is given by
\begin{equation}
\label{rel:mu}
	\mu = \hbar\omega~.
\end{equation}
Hence, as anticipated, in the relativistic embedding of the toy model all parameters are fixed in terms of the scale parameter $\omega$. With these results, the potential \eqref{qm:poschl} reads
\begin{equation}
\label{rel:poschl}
	V(r) = -\frac{3\hbar \omega}{\cosh^2(\omega r)}~.
\end{equation}

Finally, let us find the normalized wave function of the ground state. The Klein-Gordon inner product is 
\begin{equation}
\label{rel:norm}
	\left( \Psi_1, \Psi_2\right) = -i \int \rmd^3 x\, n^\mu \left( \Psi_1 D_\mu \Psi_2^\ast -\Psi_2 ^\ast D_\mu \Psi_1 \right)~, 
\end{equation}
where $D_\mu$ denotes the gauge-covariant derivative including the vector potential, and $n^\mu$ the time-like normal vector.  Explicitly,
\begin{equation}
\label{rel:Dmu}
	n^\mu D_\mu \Psi= \left(\partial_t + iV/\hbar\right) \Psi~, 
\end{equation}
and the complex conjugate for $\Psi^\ast$. Hence, because the ground state wave function is time-independent ($\epsilon_0=0$), the norm of the ground state is 
\begin{equation}
\label{rel:norm.0}
	\left( \Psi_0, \Psi_0\right) = -\frac2{\hbar} \int \rmd^3 x\, V |\Psi_0|^2~.
\end{equation}
Consider the radial wave function \eqref{qm:chi.sol}. For the ground state, the parameters are $a=-\frac12$ and $b=-\frac32$, so that the hypergeometric function in \eqref{qm:chi.sol} is just unity. Thus, together with \eqref{rel:lambda.0}, \eqref{qm:Psi.decomp} and \eqref{rel:poschl}, one finds
\begin{equation}
\label{rel:norm.0.expl}
	\left( \Psi_0, \Psi_0\right) = 6C^2 \int\limits_0^\infty \rmd x\, \frac{\sinh^2 x}{\cosh^6 x} = \frac{4}5 C^2~,
\end{equation}
from which we infer $C^2=5/4$.

%

\section{Black hole}
\label{bh}

Macroscopically, we model the graviton BEC as an anisotropic fluid. 
The local energy momentum tensor of an anisotropic fluid is of the form 
\begin{equation}
\label{bh:em.tensor}
	T^{\mu\nu} = (\varepsilon+p_\perp) u^\mu u^\nu +p_\perp g^{\mu\nu} -(p_\perp -p_\parallel) v^\mu v^\nu~,
\end{equation}
where the vectors $u^\mu$ and $v^\mu$ satisfy
\begin{equation}
\label{bh:uv.prop}
	u^\mu u_\mu = -1~,\qquad v^\mu v_\mu = 1~,\qquad u^\mu v_\mu =0~,
\end{equation}
and $\varepsilon$, $p_\perp$ and $p_\parallel$ denote the energy density and the pressures perpendicular and parallel to the space-like vector $v^\mu$, respectively. For an isotropic fluid, where $p_\perp=p_\parallel$, only the time-like $u^\mu$ is relevant. Energy-momentum conservation 
\begin{equation}
\label{bh:em.conserve}
	\nabla_\mu T^{\mu\nu} = 0
\end{equation}
imposes a relation between $\varepsilon$, $p_\perp$ and $p_\parallel$. 

For a static, spherically symmetric metric,
\begin{equation}
\label{bh:metric}
	\rmd s^2  = -f(r) \e{\gamma(r)} \rmd t^2 + \frac1{f(r)} \rmd r^2 +r^2 \rmd \Omega^2~,
\end{equation}
$u^\mu$ and $v^\mu$ are given by
\begin{equation}
\label{bh:uv}
	u^\mu = \left( f^{-1/2} \e{-\gamma/2},0,0,0\right)~,\qquad 
	v^\mu = \left( 0,f^{1/2} ,0,0\right)~.
\end{equation}
Einstein's equation 
\begin{equation}
\label{bh:einstein}
	R_{\mu\nu} -\frac12 g_{\mu\nu} R = 8\pi G T_{\mu\nu}
\end{equation}
is solved by
\begin{align}
\label{bh:f.sol}
	f(r) &= 1-\frac{2G}r M(r)~,\\
\label{bh:gamma.sol}
	\gamma'(r)&= \frac{8\pi G r}{f(r)}\left( \varepsilon+p_\parallel \right)~,
\end{align}
where the prime denotes differentiation with respect to $r$, and 
\begin{equation}
\label{bh:M}
	M(r) = 4\pi \int\limits_0^r \rmd \tilde{r}\, {\tilde{r}}^2 \varepsilon(\tilde{r})
\end{equation}
can be interpreted as the energy contained within a sphere of radius $r$.
We note that $f$ and $\gamma$ are determined by $\varepsilon$ and $p_\parallel$ alone. The perpendicular pressure $p_\perp$ follows from the energy-momentum conservation law \eqref{bh:em.conserve},
\begin{equation}
\label{bh:em.pperp}
	p_\perp = p_\parallel+\frac{r}2 \left[ p_\parallel' 
	+ \frac12 \left(\varepsilon +p_\parallel\right) \left( \frac{f'}f +\gamma' \right)\right]~.
\end{equation}

Our BEC model follows \cite{Casadio:2013hja} in using the equation of state
\begin{equation}
\label{bh:eq.state}
	\varepsilon + p_\parallel =0~,
\end{equation}
which is possible in BECs \cite{PitaevskiiStringari}, although it is not crucial in what follows. For our argument, $f(r)$ is the crucial function, and here it is sufficient to identify $\varepsilon$ with the charge density of the complex Klein-Gordon field in the ground state of our toy model. Hence, we interpret the charge of the Klein-Gordon field as a \emph{gravitational charge}. Consequently, the calculation of $M(r)$ from \eqref{bh:M} is similar to the calculation of the norm of the ground state in the previous section. Instead of \eqref{rel:norm.0.expl}, we have
\begin{equation}
\label{bh:M.expl}
	M(r) = 6 C^2 M \int\limits_0^{\omega r} \rmd x\, \frac{\sinh^2 x}{\cosh^6 x}= \frac{M}2 \tanh^3(\omega r) \left[ 5 -3 \tanh^2(\omega r) \right]~.
\end{equation} 
where $M$ is the total mass of the black hole.\footnote{It is amusing to observe that the same result would have been obtained in a purely non-relativistic toy model. In such a model, $\lambda$ would be fixed to $\lambda=3$ by the marginal binding condition, so that the ground state radial wave function would proportional to $\chi\sim \sinh(\omega r)/\cosh^3(\omega r)$. Identifying the non-relativistic toy model particle density with the energy density $\varepsilon$ would yield \eqref{bh:M.expl}. The extra factor of $1/\cosh^2(\omega r)$ in the relativistic model is provided by the potential $V$ in the Klein-Gordon norm \eqref{rel:norm.0}.} 

\begin{figure}
\begin{center}
\includegraphics[width=0.7\textwidth]{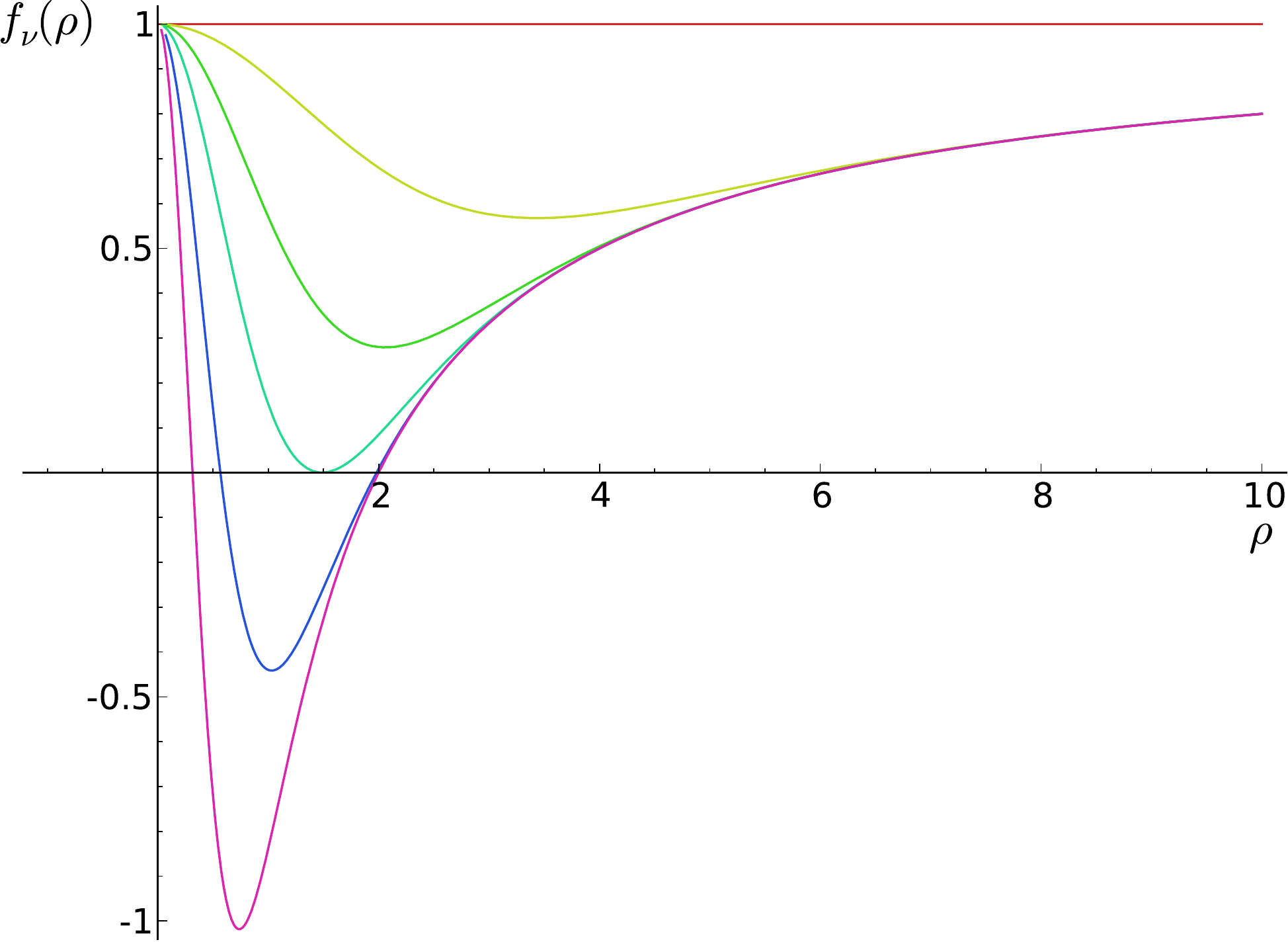}
\end{center}
\caption{Plots of $f_\nu(\rho)$ for $\nu=0,0.3,0.5,\nu_\ast,1,1.4$, from top to bottom. \label{bh:f.plot}} 
\end{figure}

To proceed, it is convenient to introduce the dimensionless variables
\begin{equation}
\label{bh:nodim.vars}
	\rho= \frac{r}{MG} = \frac{2r}{r_g}~,\qquad \nu =MG\omega~,
\end{equation}
where $r_g$ is the gravitational radius of the mass $M$. Then, with \eqref{bh:M.expl}, \eqref{bh:f.sol} becomes
\begin{equation}
\label{bh:f.nodim}
	f_\nu(\rho) = 1-\frac1{\rho} \tanh^3(\nu \rho)\left[ 5-3\tanh^2(\nu\rho)\right]~,
\end{equation}
where, for the sake of clarity, we have indicated the dependence of $f$ on the parameter $\nu$ by the subscript.
Fig.~\ref{bh:f.plot} shows plots of $f_\nu(\rho)$ for various values of $\nu$. It is evident that, for increasing values of $\nu$, the toy model black holes rapidly approach a Schwarzschild black hole.

Seen as a function of $\nu$, $f_\nu(\rho)$ is monotonically decreasing, 
\begin{equation}
\label{bh:df.dnu}
	\frac{\partial}{\partial \nu} f_\nu (\rho) = - \frac{15\sinh^2(\nu \rho)}{\cosh^6(\nu\rho)} <0~.
\end{equation} 
It follows that 
\begin{equation}
\label{bh:f.range}
	1=f_0(\rho) \geq f_\nu(\rho) \geq f_\infty(\rho) = 1-\frac{2}{\rho}~.
\end{equation}
Therefore, if an horizon exists, its radius must be smaller than the gravitational radius, $r_h\leq r_g=2MG$. 

Let us determine the condition for the existence of an horizon.
For any given $\nu$, $f_\nu(\rho)$ has a single local minimum at $\rho_\ast = x_\ast/\nu$, where $x_\ast$ is the solution of 
\begin{equation}
\label{bh:xmin}
	30 x_\ast = \sinh(2x_\ast) \left[ 4+\cosh(2x_\ast) \right]~.
\end{equation}  
Numerically, one finds
\begin{equation}
\label{bh:xmin.num}
	x_\ast = 1.03121306~.
\end{equation}
Writing
\begin{equation}
\label{bh:fmin}
	f_\nu(\rho_\ast) = 1-\frac{\nu}{\nu_\ast}~,
\end{equation}
with 
\begin{equation}
\label{bh:amin}
	\nu_\ast = \frac{x_\ast}{\tanh^3 x_\ast[5-3\tanh^2 x_\ast]} =\frac{\cosh^6x_\ast}{15\sinh^2x_\ast}= 0.69372008~,
\end{equation}
the condition for the existence of an horizon is 
\begin{equation}
\label{bh:condition}
	\nu > \nu_\ast~.
\end{equation}
The case $\nu=\nu_\ast$ is extremal.

Our final task is to show that the most natural value for $\nu$, according to the BEC picture of black holes, satisfies \eqref{bh:condition}. In our quantum-mechanical toy model, we have taken into account the marginal binding of gravitons in the condensate and minimized the ground state energy. This left us with the scale $\omega$ as the only free parameter, which determines also the effective mass of the gravitons in the ground state, $m=\mu=\hbar\omega$. Therefore, if $N$ is the mean number of gravitons in the BEC, we are led to write the quantum relation between the total mass and the scale parameter $\omega$ as 
\begin{equation}
\label{np:m.omega}
	M=N \hbar \omega~.
\end{equation} 
Together with \eqref{intro:N}, this yields the effective graviton mass and de~Broglie wavelength in Planck units
\begin{equation}
\label{np:omega.planck}
	\hbar\omega = \frac{\Mpl}{\sqrt{N}}~,\qquad \frac1{\omega} = \sqrt{N} \Lpl~,
\end{equation}
in agreement with the quantum $N$-portrait relations.
Moreover, it follows that the typical value of $\nu$ is unity. Indeed, simply rewriting $\nu$ from \eqref{bh:nodim.vars} yields
\begin{equation}
\label{np:nu}
	\nu = MG\omega = M \frac{G}{\hbar} \hbar \omega =\frac{M^2}{M^2_P} \frac{\hbar \omega}{M}=N\frac1{N}= 1~.
\end{equation}

\section{Conclusions}
\label{conc}

In this paper, we have improved upon the quantum mechanical toy model of Casadio and Orlandi for the quantum $N$-portrait of a 
black hole. Our model features the P\"oschl-Teller potential and a relativistic wave equation, such that a continuum of scattering states exists above the ground state and the scattering states possess a relativistic dispersion relation. 
The charge density of the complex Klein-Gordon field in the ground state is interpreted as a gravitational charge. It has been shown that, fixing the only parameter of the toy model to its most natural value for a BEC of gravitons, the gravitational solution which arises describes indeed a black hole. 

Besides the improvement, our model suggests some speculations about fundamental questions in general relativity. 
Take for example mass. One notion of mass in stationary space-times is the Komar mass, which can be written as the integral over contributions from the matter energy momentum tensor throughout the visible space-time, plus the masses hidden behind the horizons of black holes. Whereas the gravitational field does not contribute to the visible mass, it should carry the entire mass of the black holes, which are simply graviton BECs according to the quantum $N$-portrait. Our toy model wave function suggests that also the quantum field of gravitons carries mass, possibly in the form of a gravitational charge. Being vanishingly small for ordinary macroscopic bodies, this mass is not accounted for in classical general relativity.\footnote{A simple estimate is obtained as follows. A body of mass $M$ and size $R$ is surrounded by a coherent quantum field of $N=M^2/\Mpl^2$ gravitons, which carries a mass $m_\text{grav}=N\hbar/R=Mr_g/R$, where $r_g \sim MG$ is the gravitational radius.} This may be why in situations in which it is truly important, \ie in black holes, it is simply camouflaged by an horizon. However, there must clearly exist a cross-over regime, in which both contributions, that of matter and of the graviton field, are important. Such a situation might arise in very dense neutron stars, and it is interesting to see that there are studies of such extreme objects, which also involve Bose-Einstein condensation \cite{Chavanis:2011cz}.

\section*{Acknowledgements}
We are grateful to Roberto Casadio and Alessio Orlandi for stimulating discussions. This work was supported by the European Science and Technology (COST) action ``The String Theory Universe'', the European Science Foundation's HoloGrav network and the INFN research initiative ``String Theory and Fundamental Interactions''.

\bibliography{}
%
\end{document}